**The Structure of Online Social Networks Mirror Those in the Offline World**

**R.I.M. Dunbar [1,*], Valerio Arnaboldi [1,2], Marco Conti [2], Andrea Passarella [2]**

1) Department of Experimental Psychology, University of Oxford, Oxford, UK

2) Institute of Informatics and Telematics of CNR, Pisa, Italy

*Correspondence*:

R.I.M. Dunbar

Department of Experimental Psychology, University of Oxford, South Parks Rd, Oxford OX1 3UD, UK  [robin.dunbar@psy.ox.ac.uk]

**Abstract**

We use data on frequencies of bi-directional posts to define edges (or relationships) in two Facebook datasets and a Twitter dataset and use these to create ego-centric social networks. We explore the internal structure of these networks to determine whether they have the same kind of layered structure as has been found in offline face-to-face networks (which have a distinctively scaled structure with successively inclusive layers at 5, 15, 50 and 150 alters). The two Facebook datasets are best described by a four-layer structure and the Twitter dataset by a five-layer structure. The absolute sizes of these layers and the mean frequencies of contact with alters within each layer match very closely the observed values from offline networks. In addition, all three datasets reveal the existence of an innermost network layer at ~1.5 alters. Our analyses thus confirm the existence of the layered structure of ego-centric social networks with a very much larger sample (in total, >185,000 egos) than those previously used to describe them, as well as identifying the existence of an additional network layer whose existence was only hypothesised in offline social networks. In addition, our analyses indicate that online communities have very similar structural characteristics to offline face-to-face networks.

*Key Words:* ego-centric social networks, network scaling, network structure, Facebook, Twitter,

1. **Introduction**

The growth of digital communication (and, in particular, social networking sites) over the past decade has raised fundamental questions about the constraints that exist over both the size and the pattern of social relationships. In one sense, the implicit promise of the new technologies was that they would open up the vista of a social world that was intrinsically unlimited in size. This becomes of particular interest in the light of the finding that there appears to be a cognitive limit on the size of natural face-to-face social networks (Dunbar 1993, Roberts et al. 2009, Sutcliffe et al. 2012). This limit is thought to arise out of a combination of a cognitive constraint and a time constraint.

The central cognitive constraint, known broadly as the social brain hypothesis, is based on the observation that, in primates, the typical size of social groups correlates closely with the size of the neocortex (Dunbar 1992), and in particular with the more frontal units of the neocortex (Joffe and Dunbar 1997, Dunbar 2011). This seems to imply that in some way the information-processing capacity of the brain limits the number of relationships that individuals of a particular species can manage, thus limiting the size of groups because they become unstable and prone to fission when they exceed this size. Species with larger (frontal) neocortices manage to maintain coherence in larger groups than those with smaller neocortices. This proposal has since been given considerable support by evidence from a series of neuroimaging studies which have shown, for both humans (Lewis et al. 2011, Powell et al. 2012, 2014, Kanai et al. 2012) and monkeys (Sallet et al. 2013), that within-species variation in social network size correlates with the volumes of particular brain regions at the level of the individual. Powell et al. (2012) showed that, at least in humans, this relationship is mediated by mentalising competences. Mentalising competences (most commonly associated with theory of mind or mindreading, the ability to understand another individual's mental state) form a natural recursion running from first order (the state of self-consciousness) through second order (formal theory of mind) to fifth order in normal human adults, with a range in adults of around fourth to seventh order (Kinderman et al. 2008, Stiller and Dunbar 2007). Powell et al. (2012) were able to show that there was a causal relationship in which the volume of the orbitofrontal cortex determined mentalising skills, and mentalising skills determined network size.

In addition, however, there is also evidence to suggest that time imposes a constraint. Time becomes important because it seems that the strength of a relationship is determined by how much time two individuals spend together. In humans, self-rated estimates of the emotional closeness for dyadic relationships (using a simple 0-10 analogue scale) correlate closely with the frequency of contact (Roberts and Dunbar 2011, Arnaboldi et al. 2013), and these in turn correlate with willingness to behave altruistically towards the alter in question (Curry et al. 2013). Similar findings have been reported for monkeys (Dunbar 2012). One reason for this is that time (and maybe social/emotional capital) is limited (Miritiello et al. 2013) and individuals are forced to choose between investing their time and/or emotional capital thickly among a small number of alters or thinly among a larger number. Pollet et al. (2011a), for example, found that although extroverts typically had more individuals in their social networks than introverts, their average self-rated emotional closeness to these individuals was significantly lower. Similarly, Roberts and Dunbar (2011) found that individuals who had larger social networks distributed their available social capital (as indexed by their self-reported emotional closeness) more thinly than those who had smaller networks.

Three studies of digital datasets have sought to determine whether social networks online are also limited in size, and if so to what size. Pollet et al. (2011b) examined the offline social network of heavy and casual users of internet social networking sites, and found that they did not differ. Gonçalves et al. (2011) downloaded traffic among the followers of individual Twitter accounts and, using a criterion of reciprocated exchanges to identify meaningful relationships, concluded that Twitter communities typically averaged between 100-200 individuals. Similarly, in an analysis of email traffic among physicists, Hearter et al. (2012) found, using a similar definition to identify relationships, that there was a marked downturn in the rate at which additional members were acquired once communities exceeded 200 individuals.

Individuals do not, however, distribute their social effort (whether measured by time or by self-rated emotional closeness) evenly among the alters in their networks. Indeed, there is considerable evidence to show that, within natural social networks, individual alters can be ranked in order of declining investment by ego (e.g. Saramäki et al. 2014) and that these rankings fall into a natural series of layers with a scaling ratio of ~3 that yields breakpoints at around 5, 15, 50 and 150 alters (Zhou et al. 2005, Hamilton et al. 2007). These layers correspond to marked differences in both the frequency of contact with alters and in rated emotional closeness (Roberts et al. 2009, Sutcliffe et al. 2012), seemingly reflecting a combination of temporal and cognitive constraints that give rise to the layered structure of networks.

We here combine these two sets of findings and ask whether, given that internet-based communication might be expected to by-pass at least some of the time constraints that limit face-to-face networks, online social networks nonetheless still exhibit the same kind of structuring. For these purposes, we examine three online datasets, two of them culled from Facebook (Viswanath et al. 2009, Wilson et al. 2012) and a Twitter dataset specially downloaded for the purpose (Arnaboldi et al. 2013b). In each case, we use specific algorithms to search for patterns in the data so as to determine, first, whether a layered structure similar to the one found in offline ego networks is present in the reciprocated traffic data collected form online environments, and then, if so, to identify the sizes of these layers.

## 2. Methods

### 2.1. *Facebook dataset #1*

Facebook dataset #1 was obtained before 2009 when the default privacy settings allowed users inside the same regional network to have full access to each others' personal data (Wilson et al. 2012). This dataset has been widely used for social network analysis (see for example Arnaboldi et al. 2012). The dataset covers the time span from the start of Facebook in September 2004 until April 2008 (it is publicly available for research and can be accessed at http://current.cs.ucsb.edu/facebook/, "Anonymous regional network A"). As explained in (Wilson et al. 2012), the dataset represents only a subsample of the original Facebook regional network, in terms of downloaded Facebook profiles (~56%) and their Facebook friendships (~37%). Although other analyses on Facebook ego network structure have been conducted using this dataset (Arnaboldi et al. 2012), here we will improve existing results through a more refined analysis of the dataset, obtaining more accurate results about the size and the composition of ego network layers.

The dataset was downloaded using a crawling agent that obtained the complete public profile information (including personal information and the list of Facebook friends), and the Facebook wall data of a set of users in a large regional network of Facebook. The agent followed the friendship links to obtain a large connected component of the regional network. The 44% of profiles in the regional network that was not been downloaded were profiles with restrictive privacy settings or users disconnected from the giant component. Despite the high number of missing profiles, some of their data is still present in the dataset. In fact, if a public profile of a user A was connected to a non-public profile B, the posts sent from B to A were still visible in A's Facebook wall. Moreover, B would appear in the friend list of A. Therefore, information exchanged on missing links *from non-public* profiles *to public* profiles is still available. We miss information related to posts (i) *from public* profiles (node A in our example) *to non-public* profiles (node B) and (ii) between non-public profiles. We discuss below how we estimate traffic related to (i). As for (ii), the amount of data collected for non-public profiles is usually lower than that of public profiles since their communication traces appear only indirectly inside the walls of other public users. For this reason, most private profiles appear as users with low Facebook usage, which we discard in our analysis. Given this, we argue that missing information about their mutual interaction is not particularly problematic for our purposes. Hence, we reasonably assume that, despite not containing all the possible communication records between users in the regional network, the dataset is still a valid representation of Facebook social network for the purpose of ego network analysis.

We managed to partly reconstruct missing information in respect of point (i) above, as follows. We cannot tell from the dataset itself which profiles are public and which are not because, for a given friendship relationship, the dataset only reports the number of (undirected) interactions (posts or photo comments) that occurred, and not the properties of the profiles of the users involved, or the detailed interaction log. Therefore, we do not know for which links in the dataset we are missing interactions in one of the two directions. The only information we have is the percentage of non-public profiles, i.e. 44%. For this reason, we have selected randomly 44% of nodes, and assumed that those are associated with the non-public profiles[1]. We have doubled the number of interactions on all the links of the ego networks of those nodes. This corresponds to assuming that these relationships are perfectly bi-directional, and the (unknown) amount of interaction from public to non-public profiles is the same as the (known) amount of interaction in the opposite direction. We can expect that this process makes our results for internal layers accurate and less precise for external layers, for the following reasons. First, it is known that bi-directionality becomes stronger and stronger as relationships become more and more intimate (R. Dunbar, unpublished data). Therefore, doubling interactions on strong relationships is very reasonable, while it is less accurate for weak ties. In addition, this process does not modify interactions over links for which we have no interactions in the dataset. Hence, after adjusting the amount of interactions there may still be some relationships for which we incorrectly consider no interactions, i.e. relationships for which *real* interactions have occurred only in the direction from a public to a non-public profile. Such strongly asymmetrical relationships are typically known to belong to the most external layers of the ego networks. The net expected effect is, therefore, that the size of internal layers is precise, while that of external layers may be underestimated.Facebook dataset #1 consists of 3+ million nodes and 23+ million edges (social links identified by cross-postings), with an edge representing a Facebook friendship. The dataset provides an approximate measure of time in so far as the data are coded into four time periods (postings or contacts within the last month, last six months, last year and the entire duration of the link), which we use to identify four time-based windows $w_k$, ($k \in \{1, 2, 3, 4\}$) with $w_1$ indicating the time interval between the download and one month before the download (last month), $w_2$ between one month before the download and six months before the download (last six months) and so on.

---

[1] We assume that the amount of non-public nodes without any connection to public nodes is negligible.

We define as "active" all the relationships that have at least one interaction in any of the windows $w_k$. For each link, we use the difference between the number of interactions made in the different temporal windows to compute contact frequency, and we interpret this as an estimate of the intimacy of the relationship. A complete description of the methodology we used to obtain the contact frequency of the relationships in the dataset can be found in Arnaboldi et al. (2012). The complementary cumulative distribution function (CCDF) of the contact frequency is depicted in Figure 1 for the three datasets. The distributions have a long tail shape. This means that the contact frequency is low for most of the relationships, but there are a few relationship with very high levels of interactions. This type of distribution is typical in social networks.

For the analysis we consider only egos with an average of more than 10 interactions per month, thus selecting "socially active people" since they are particularly relevant for our analysis, and discard inactive profiles. The final dataset contain 130,338 egos with 5,289,910 active edges (i.e. friendships with at least one interaction). Note that, to extract ego networks from the datasets, we first create a series of sets each of which contains all the social relationships of a user. The CCDF of the size of ego networks in each dataset is depicted in Figure 2. It is worth noting that, for the majority of ego networks, the size is lower than 100. This means that even though people can potentially add up to 5,000 friends in Facebook, they communicate only with a small subset of them.

We further refine the dataset by selecting, for each ego network, only the set of relationships with contact frequency higher than one message per year. This is to avoid considering people in whom the ego does not invest some minimum amount of time and cognitive resources. We choose the limit as one message per year in accordance with the definition of active network in offline ego networks (Hill & Dunbar 2003). In this way, we can avoid considering ego network layers external to the active network, which still lack a precise definition, and whose properties are not completely known. Figure 3 depicts the CCDF of the active network size of the ego networks in each dataset under this added constraint. As it can be noted by comparing Figures 2 3, the selection of social relationships with more than one message per year does not represent a substantial change in the size of the ego networks, but, on the other hand, it helps to eliminate relationships external to the active network representing noise for our analysis since they are not associated to a meaningful level of effort in terms of time and cognitive resources spent by ego to maintain the relationships over time.

*2.2.   Facebook dataset #2*

Facebook dataset #2, similarly to Facebook dataset #1, was downloaded in 2009 exploiting the Facebook regional network feature. It represents the Facebook regional network of New Orleans and it has been obtained through a crawling agent similar to the one created for downloading Facebook dataset #1 (Viswanath et al. 2009). Compared to the first dataset, Facebook dataset #2 represents a smaller regional network (90,269 nodes and 3,646,662 social links), but the data it contains are much more detailed. Specifically, for each public profile visited by the crawler, the dataset reports the list of its Facebook friends and the list of wall posts received by the user from her friends, with the timestamp indicating the time at which the interaction occurred. In contrast to Facebook dataset #1, where it was not possible to identify the set of public profiles, here we know exactly which public profile has been visited by the crawling agent, and we can perform a more precise analysis, taking only public profiles as the egos for our ego network analysis. For these profiles, we know the exact size of the ego network (as we know the entire set of friends), and we can reconstruct almost entirely information about interactions, as described below. After selecting public profiles and all their social interactions, we obtain a dataset containing 60,290 nodes (egos) with a total of 1,545,686 social links. The data collected for each ego represents her Facebook wall. For this reason, they contain only the communications received by the users from her friends. For friends with public profiles, we can complement the information available on the ego's wall and reconstruct the exact number of mutual interactions by analysing the friends' wall, where posts and photo comments made by ego are available. As in the case of Facebook dataset #1, we doubled the number of interactions available on the ego's wall for friends with private profiles (which may result in approximations primarily in external layers of ego networks, as discussed above).

Even though Facebook dataset #2 contains more accurate information than Facebook dataset #1, the larger size of the latter makes it a more significant sample of the entire Facebook network. The main purpose of the analysis on Facebook dataset #2 is to validate results obtained from dataset #1. To this end, we used dataset #2 also to validate the reconstruction methodology applied to dataset #1, as follows. In dataset #2, 58% of the nodes are non-public. We apply the same reconstruction methodology used in dataset #1, by sampling 58% of nodes from the entire dataset, and doubling the interactions for all their links. We then repeat the analysis to characterise ego network structures, and compare the results obtained with the two different reconstruction methodologies applied to dataset #2.

We calculate the frequency of contact between users in Facebook dataset #2 as the number of interactions between them divided by the duration of their relationship, estimated as the time since their first contact, considered from the time of the download. In contrast to Facebook dataset #1, here we have precise information about the duration of the relationships. The CCDF of the contact frequency in the dataset is depicted in Figure 4. The distribution is very similar to the one of dataset #1 (in Figure 1), even though it shows a slightly shorter tail. The presence of a shorter tail could be ascribed to the smaller size of the dataset compared to Facebook dataset #1. In fact, since relationships with very high contact frequency are rare in the whole Facebook network, a smaller sample has a lower probability to contain high contact frequency.

Also in this case, we select "socially relevant users", taking into consideration only users who had at least 10 interactions per month. After this pre-processing the dataset contains 5,761 egos and 107,029 social relationships. The CCDF of the size of ego networks in the dataset and of the active network size (after selecting only relationships with contact frequency of at least one message per year) are depicted in Figures 5 and 6 respectively. Also for these figures, the distributions are similar to those of dataset #1, but with a slightly shorter tail, that could be explained (similarly to the differences in terms of contact frequency) by the smaller size of the dataset.

## 2.3. Twitter dataset

We downloaded a sample of 303,902 Twitter user profiles by crawling Twitter in November 2012, using the crawler agent described in Arnaboldi et al. (2013b), which follows links between users to build a network of connected profiles.

Twitter followers can interact with each other through the *mention* and *reply* functions that allow direct communication between users. Besides direct communication, all the tweets are automatically broadcast to all the users' followers. Tweets can be *retweeted* or forwarded by users to all their followers.

For each profile visited by the crawler, we extract all the Twitter replies sent to other users to calculate the frequency of contact with its social contacts. Using replies allows us to capture the intentionality in the communications, since a reply represents a direct message sent from the considered user to one of her contacts. As with mentions, replies allow users to "mention" other users including a reference to their username in the tweet. Mentions are often used to send a tweet to multiple recipients, whereas replies usually include only one mentioned profile and represent for this reason a better proxy for the emotional investment of the user in the respective relationships. We calculate the frequency of contact of a relationship $r$ in Twitter, involving two users $u_1$ and $u_2$, as follows:

$$f(r) = \frac{N_{rep}(r)}{d(r)}$$

where $N_{rep}(r)$ is the number of replies sent by $u_1$ to $u_2$ and $d(r)$ is the duration of the relationship $r$ between them, calculated as the time since first interaction (mention or reply). As with Facebook dataset #2, we are able to use the exact value of the duration of the relationship, since we have access to the timestamp of the tweets sent by the users.

To avoid including Twitter users that are not human (companies, institutions, bots, etc), we filter the data identifying user profiles that have recognisable human characteristics (as explained in detail in Arnaboldi et al. 2013c). The CCDF of the contact frequency of relationships of human users is depicted in Figure 7. The figure shows a higher level of interaction in Twitter than in Facebook, even though the long tailed shape of the distributions is qualitatively similar.

As with the Facebook datasets, we selected only accounts that had an average of more than 10 interactions per month. The final dataset contains 60,790 egos and 5,323,195 social relationships. The CCDF of the size of the ego networks in the dataset, and the size of the active networks are depicted in Figures 8 and 9, respectively. The CCDFs show longer tails than in Facebook. This indicates that in Twitter there are users with larger ego networks than in Facebook. Nevertheless, similarly to Facebook, more than 90% of the Twitter users have less than 100 relationships. This means that only a very limited number of users seem to have larger ego networks than in Facebook but, for the majority of the users in Twitter, the ego network sizes are comparable with those found in the other datasets.

*2.3 Analysis*

We use two different clustering techniques (k-means, a *partitioning clustering* technique, and DBSCAN, a *density-based clustering* technique) on the frequency of contact of each ego network to search for a layered structure. Partitioning clustering algorithms start with a set of objects and divide the data space into *k* clusters so that the objects inside a cluster are more similar to each other than objects in other clusters. For each ego network, we order alters in a one dimensional space by contact frequency with the ego, and search for clusters in this one dimensional space using the technique described in Wang & Song (2011).

Density-based clustering algorithms are able to identify clusters in a space of objects with areas with different densities (Kriegel et al. 2011).

The *k*-means approach involves partitioning the data space into *k* different clusters of objects, so that the sum of squared Euclidean distances between the centre of each cluster and the individual objects inside that cluster is minimised. The goodness-of-fit of *k*-means algorithm is often expressed in terms of variance explained, defined as follows:

$$VAR_{exp} = \frac{SS_{TOT} - \sum_{i=1}^{k} SS_i}{SS_{TOT}}$$

where $SS_{TOT}$ is the total sum of squares in the data space and $SS_i$ is the within sum of square of the i[th] cluster. $VAR_{exp}$ is analogous to the conventional coefficient of determination $R^2$, which ranges between 0 and 1.

We apply *k*-means in two different ways. On the one hand, we want to find the typical number of clusters in the ego networks, as we want to verify if Facebook and Twitter ego networks show a layered structure with a number of layers similar to the one found in offline social networks. To do so, we apply *k*-means to each ego network with different values of *k*. However, since $VAR_{exp}$ will always be maximised when *k* is equal to the number of objects in the data space, we need an algorithm to avoid this over-fitting problem in order to discover the optimal number of clusters in our ego networks. To do this, we calculate the Akaike Information Criterion index (AIC) of the model for each *k*-mean configuration, and, by varying *k* from 1 to 20, take the value of *k* that minimises the value of AIC. This value is the optimal number of clusters $k^*$ for the data. This is a standard approach for the analysis of the optimal number of clusters in a data space (Akaike 1974).

The AIC is a measure representing the relative quality of the clustering configurations. It is useful to find the optimal number of clusters in the data, but it does not indicate a direct measure of the goodness of a single configuration since it only provides a ranking between possible configurations. For this reason, even if we are able to identify the optimal number of clusters in the contact frequency of an ego network by using the AIC, we do not know if the data are effectively centred on the centroids of the clusters or not. To measure how well the data are clustered, we calculate the *silhouette* statistics for each optimal configuration. The silhouette of a single data point *x, s(x)*, measures the extent to which *x* is appropriately assigned to its cluster. Specifically, given *a(x)*, the average distance between *x* and all the other points in the same cluster, and *b(x)*, the minimum amongst the average distances between *x* and elements in each of the other clusters, *s(x)* is defined as follows:

$$s(x) = \frac{a(x) - b(x)}{\max\{a(x), b(x)\}}$$

The average value of *s(x)* for all *x* in the data space indicates the appropriateness of the clustering configuration. The value of *s(x)* ranges between -1 and 1. Values close to -1 indicate that the points could have been assigned to the wrong clusters. On the other hand, values close to 0 indicate that the data are poorly clustered and thus uniformly distributed around centroids. In this case the clustering configuration could be sub-optimal, or the data could not be naturally grouped into clusters. Lastly, values of *s(x)* close to 1 indicate that the data is appropriately clustered, and densely distributed around centroids. If the best clustering configuration identified through the AIC is associated to a value of s(x) close to 1, the data are naturally divided into groups that are effectively matched by the obtained clusters.

After the identification of the optimal number of clusters $k^*$, we verify the appropriateness of the clustering configuration through the silhouette statistic, and then apply k-means with $k=k^*$, and study the average size and the minimum contact frequency of each cluster obtained (i.e. the minimum contact frequency required for an alter in order to be part of the cluster).

The results obtained with *k*-means may be affected by the presence of noisy data (i.e. points in the data space with a very low density compared to the other points around them). Noise can affect a *k*-means analysis in two different ways: (i) the presence of noisy points between two adjacent clusters might cause the algorithm to treat them all as a single cluster instead of two (the so called "single link effect": Kriegel et al. 2011), while (ii) the presence of a large number of noisy points in the data set could lead to the detection of more clusters than really exist. To ensure that noisy points do not adversely influence the outcome, we compare the results with the DBSCAN density-based clustering algorithm (Ester et al. 1996). DBSCAN defines two parameters, $\varepsilon$ and *MinPts*, and any object with more than *MinPts* neighbours within a distance $\varepsilon$ is defined as a core object. A cluster is made up by a group of core objects (adjacent elements separated by less than $\varepsilon$) and by the "border objects" of the cluster (termed 'non-core objects') linked to a core object at a distance less than $\varepsilon$. For a more formal definition of density based clusters, see Ester et al. (1996). Points with less than *MinPts* neighbours within a distance $\varepsilon$ that are not border objects are considered noise by DBSCAN, and they are excluded from the clusters. We iterate DBSCAN decreasing the value of $\varepsilon$ until we find a number of clusters equal to the number of clusters obtained by *k*-means. Hence, by comparing the results of *k*-means and DBSCAN in terms of cluster size, we can verify that the former are valid and not influenced by noisy points. To allow noisy data to be identified by the iterative DBSCAN procedure, we set the parameter *MinPts*=2. In this way, isolated points are excluded from the clusters.

## 3. Results

Figure 10 plots, for each dataset, the distribution of the optimal number of clusters $k^*$ of the ego networks of each dataset, obtained by iteratively applying *k*-means. The distributions show a marked peak around $k^*=4$ for all the datasets. For Facebook dataset #1, the ego networks have an average optimal number of clusters equal to 4.35 (with median 4), and Facebook dataset #2 has an average optimal number of clusters of 4.10 (with median 4).

Despite a clear mode at 4, the ego networks in the Twitter dataset have an average optimal number of clusters equal to 6.60 (with median 5) due to the long tail to the right.

The average silhouette value for the best configurations associated with the optimal number of clusters for each ego network is 0.670 for Facebook dataset #1, 0.678 for Facebook dataset #2, and 0.674 for Twitter. These values indicate that the data are appropriately clustered, and that the identified clusters are not fictitious.

To be able to study the average size and composition of the ego network layers in the datasets we must apply *k*-means fixing a value of *k* for each dataset. Since the distribution of the values of $k^*$ depicted in Figure 1 have a modal value around 4, that for Facebook is also very close to both the average value and the median, and for Twitter the median is 5, we apply *k*-means with *k*=4 on the Facebook ego networks and with *k*={4,5} on the Twitter ego networks. Then, we calculate the average size of each layer. We report the respective values for each layer in Table 1. Note that the layer sizes obtained by this method are the numbers of individuals in each annulus, not the cumulative number of individuals in successive layers as defined in the literature. To obtain the layers equivalent to those defined in the literature, we nest the clusters, obtaining a series of concentric layers. We check the sizes of these layers using the density-based clustering algorithm DBSCAN to control for noisy points, and the results, as reported in Table 1, are very similar. Specifically, in all the datasets, the last layer found by DBSCAN differs for a maximum of about 3 elements from the results of *k*-means[2]. For all the datasets, the results of the remaining layers for *k*-means and DBSCAN are very similar, confirming that *k*-means, despite its simplicity, is able to identify the correct clusters in the ego networks data.

---

[2] This means that, on average, only 3 alters have contact frequency with the ego significantly far from the other values in the ego network, and are thus considered noise by DBSCAN. The difference in terms of size of the innermost layer could be due to the presence of a minimum of 2 elements in each cluster required by DBSCAN. In fact, in the results found by *k*-means, this layer is often composed of a single element, and in DBSCAN is forced to have at least 2 members.

Figures 2-4 plot the CCDF distributions for contact frequency for each dataset separately.

We denote as layer 1 to 4 the layers in the conventional ego network model obtained with offline datasets. As we discuss hereafter, we have found an additional internal layer that has not previously been identified, which we denote as layer 0. The scaling ratio between layers are very close to 3 for all the $k$-means configurations and all the datasets. For comparison, Table 1 also gives the characteristic sizes of each of these layers in offline ego-centric personal social networks, as determined from face-to-face contacts, for which Zhou et al. (2005) found a scaling ratio of ~3.2. Note that for the Twitter dataset, the results with $k$=5 match the offline layer sizes much better than those for $k$=4 (which tends to combine two of the middle layers) for the results found by $k$-means. On the other hand, the results found by DBSCAN seem to indicate that the best configuration is the one with $k$=4.

Table 1 also gives, for the three datasets, the average value of the minimum contact frequency of the layers (i.e. the lower frequency of the layers, indicating the boundaries between them). The frequencies are expressed in number of posts sent by the ego per year. These suggest that in Facebook individual alters are contacted approximately at least every five days for layer 0, at least every twelve days for layer 1, at least once a month for layer 2, and at least once every six months for layer 3. These values are compatible with those obtained for face-to-face networks (Sutcliffe et al. 2012). In the Twitter dataset the contact frequencies are higher, with alters contacted approximately at least once every one/two days in layer 0, at least every three days in layer 1, at least once a week in layer 2, at least once a month in layer 3, and at least two/three times a year in layer 4. This can be attributed to the specificity of this OSN platform that is explicitly designed for the exchange of short and frequent messages between users (micro-blogging).Bearing in mind this difference between Facebook and Twitter, we match (as reported in Table 1) the layers we have found in online ego networks with those in face-to-face networks, on the basis of their contact frequency. To be able to compare the results found in Twitter with the other ones, we consider that in Twitter the contact frequency is between two and four times higher than in Facebook, as can be also noted in the results reported in Table 1, especially for the innermost layers.

In Table 1, we also report the results of the analysis on Facebook dataset #2 aimed at validating the procedure used on Facebook dataset #1 to reconstruct missing data about the communication from public to private user profiles. The size of the ego network layers in Facebook dataset #2 and in the same dataset pre-processed with the reconstruction procedure used for dataset #1 are really similar. The minimum contact frequencies of the layers, when using the reconstruction procedure of dataset #1, are slightly higher than the correct one, and similar to those found in Facebook dataset #1. This seems to indicate that the results of Facebook dataset #1, in terms of minimum contact frequency, could be slightly overestimated, and should be more similar to those found for Facebook dataset #2.

## 4. Discussion

Our analyses of three different online datasets confirm the layered structure found in offline face-to-face social networks. For all the online datasets, the scaling ratio for the various layers identified by the analyses, and the respective sizes of these layers, are extremely close to those observed in offline networks (Hill and Dunbar 2003, Zhou et al. 2005, Hamilton et al. 2007). These layers have previously been identified only from samples of quite modest size (grouping levels in small scale societies, Christmas card distribution lists: all N<<1000). The present analyses provide us with strong evidence both for the existence of these layers and for their relative sizes and scaling ratios based on (a) very large sample sizes (>130,000, ~6000 and ~61,000 egos, respectively) from (b) three very different sources.

The sizes of the entire ego networks for the three datasets are smaller than the total size of conventional offline egocentric networks (typically about 150 alters: Hill and Dunbar 2003, Roberts et al. 2009). This is true especially for the Facebook datasets, where the most external ego network layer is completely missing. This is, perhaps, not too surprising, since the outermost (150) layer in offline networks corresponds to people who are contacted only about once a year. At least as far as the Facebook datasets are concerned, early users (remember that datasets were collected in 2009, when Facebook was still new and yet to start its exponentially increasing diffusion) were not forced into 'friending' complete strangers and, instead, typically only sought out people they knew. Moreover, as discussed earlier, the information we have about weak relationships in Facebook datasets could be not enough to identify all the relationships with very low contact frequency.

In Table 1, we match all the layers obtained from online communications datasets and those identified offline, according to their minimum frequency of contact. The contact frequency of the ego network layers in the two Facebook datasets is significantly lower than those in the corresponding layers in the Twitter dataset. This is not surprising since at the time of the download (2009) users were less active on Facebook than now. Moreover, tweets are very short messages (i.e. up to 140 characters), and the emotional investment of a tweet is likely to be lower than an interaction in Facebook or face-to-face. Despite this, the data on contact frequencies yielded by the online data are surprisingly similar to the face-to-face contact frequencies observed in offline networks. Although the Twitter contact rates are, generally, somewhat higher than both the Facebook and face-to-face contact rates (perhaps reflecting a lower time and emotional investment), nonetheless they are broadly similar, being only about double the frequency of the latter, and in similar ratios across the layers.

Aside from confirming the size and scaling ratios for the conventional first three layers of social networks (those associated, cumulatively, with 5, 15 and 50 individuals) for the Facebook datasets, and also the outermost layer (associated with 150 individuals) for the Twitter dataset, the three online datasets also identify an entirely new layer that was not visible from face-to-face communication data (layer 0 in Table 1). This is an innermost layer at ~1.5 individuals, scaling perfectly with the layers outside it. The layer is visible in all the datasets. It is clear that this innermost layer has special relevance to egos since they contact these individuals at very high frequencies (on average at least once every five days in Facebook and every other day in Twitter), indicating a very high emotional investment.

This tendency for individuals to have one or two intensely intimate friends is evident in other smaller datasets where contact frequencies have been plotted in rank order (e.g. Saramäki et al. 2014). However, it is clear from even the data shown in Saramäki et al. (2014) that not everyone has such an innermost layer, perhaps explaining the rather odd fact that the scaling ratio itself predicts a decimal value for this layer. It is also possible, however, that this reflects a gender difference in attachment to intimates, such that men have 0-1 and women 1-2. Unfortunately, we do not know the gender identities of the egos in either dataset, so we cannot test whether or not this is so. However, the fact that the three large-scale datasets we have analysed identify this innermost layer suggests that it really is a robust phenomenon, and would merit closer attention.

Quite remarkably, the mean rates of contact in each layer are extremely close, especially for the Facebook datasets, to those found in (and, indeed, used to define: Dunbar and Spoors 1995) the different layers in egocentric offline personal social networks (Sutcliffe et al. 2012). This suggests that the online environments may be mapping quite closely onto everyday offline networks, or that individuals who inhabit online environments on a regular basis begin to include individuals that they have met online into their general personal social network, treating the different modes of communication as essentially the same. This, of course, has important implications for both the design and promotion of online social environments. However, our present concern is with the sociological similarities between online and offline environments, as implied by these data.

**Acknowledgments**

RD's research is supported by a European Research Council Advanced grant.

**Table 1. Mean (±SD) number of alters in each of the four layers identified by a *k*-means analysis with *k*={3,4,5} for Facebook and Twitter datasets.**

| Layer | 0 | 1 | 2 | 3 | 4 |
|---|---|---|---|---|---|
| **Offline networks*** | | | | | |
| Number of alters[#] | ? | 5 | 15 | 50 | 150 |
| Contact frequency[§] | ? | 48 | 12 | 2 | 1 |
| **Facebook dataset #1, k=4** | | | | | |
| Number of alters[#] | 1.68±0.01 | 5.28±0.02 | 14.92±0.06 | 40.93±0.20 | - |
| Contact frequency[§] | 77.36±0.77 | 30.28±0.24 | 11.15±0.07 | 2.53±0.01 | - |
| DBSCAN size | 2.87±0.01 | 7.34±0.03 | 18.86±0.09 | 37.53±0.20 | - |
| **Facebook dataset #2, k=4** | | | | | |
| Number of alters[#] | 1.53±0.03 | 4.34±0.09 | 10.72±0.23 | 26.99±0.61 | - |
| Contact frequency[§] | 58.54±2.62 | 22.19±0.74 | 7.93±0.23 | 1.37±0.04 | - |
| DBSCAN size | 2.57±0.05 | 5.86±0.10 | 10.65±0.20 | 27.02±0.69 | - |
| **Twitter dataset, k=4** | | | | | |
| Number of alters[#] | 1.87±0.03 | 6.54±0.09 | 21.09±0.27 | | 88.31±0.87 |
| Contact frequency[§] | 259.53±4.04 | 93.03±1.31 | 26.92±0.38 | | 2.54±0.02 |
| DBSCAN size | 2.79±0.03 | 6.86±0.11 | 14.24±0.21 | | 77.72±1.15 |
| **Twitter dataset, k=5** | | | | | |
| Number of alters[#] | 1.55±0.02 | 4.52±0.06 | 11.17±0.15 | 28.28±0.32 | 88.31±0.87 |
| Contact frequency[§] | 276.63±4.06 | 113.12±1.49 | 49.63±0.66 | 16.89±0.21 | 2.54±0.02 |
| DBSCAN size | 2.59±0.03 | 5.99±0.07 | 10.71±0.13 | 19.51±0.28 | 85.35±1.15 |
| **Facebook dataset #2 for validation, k=4** | | | | | |
| Number of alters[#] | 1.50±0.03 | 4.22±0.07 | 10.13±0.17 | 25.02±0.48 | - |
| Contact frequency[§] | 72.95±2.62 | 27.90±0.73 | 10.35±0.24 | 1.83±0.05 | - |

*Egocentric personal social networks determined from face-to-face contacts (Zhou et al. 2005; Roberts et al. 2007).
[#]Cumulative number of alters across layers.
[§]Average frequency of reciprocated posts per alter per year.

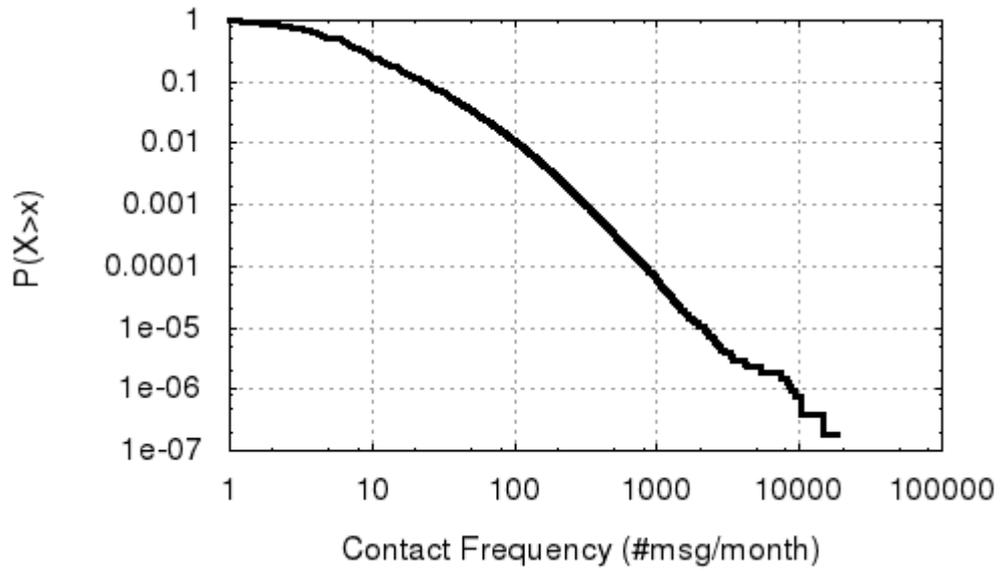

**Figure 1. CCDF of the contact frequency for relationships in Facebook dataset #1**

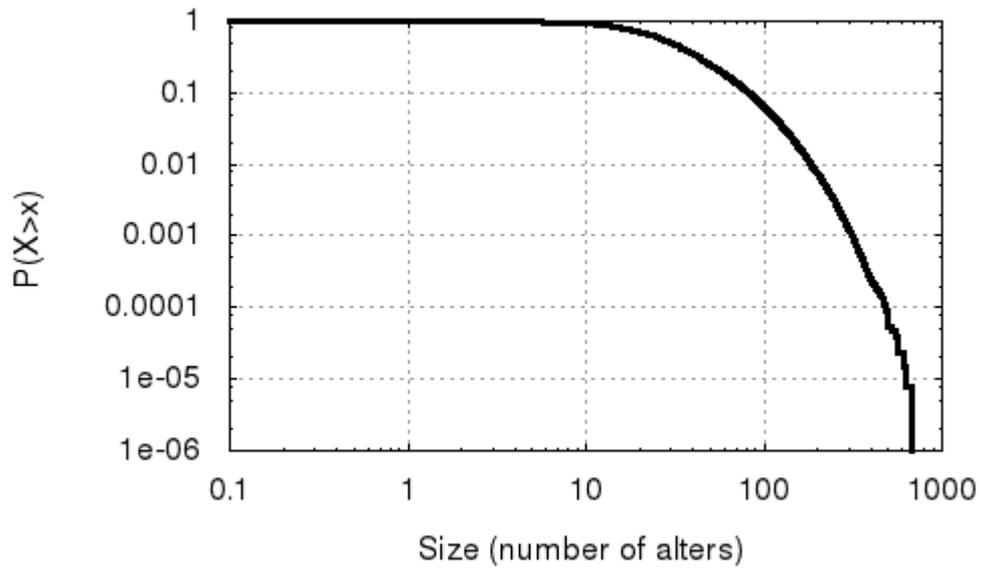

**Figure 2. CCDF of the size of ego networks for relationships in Facebook dataset #1**

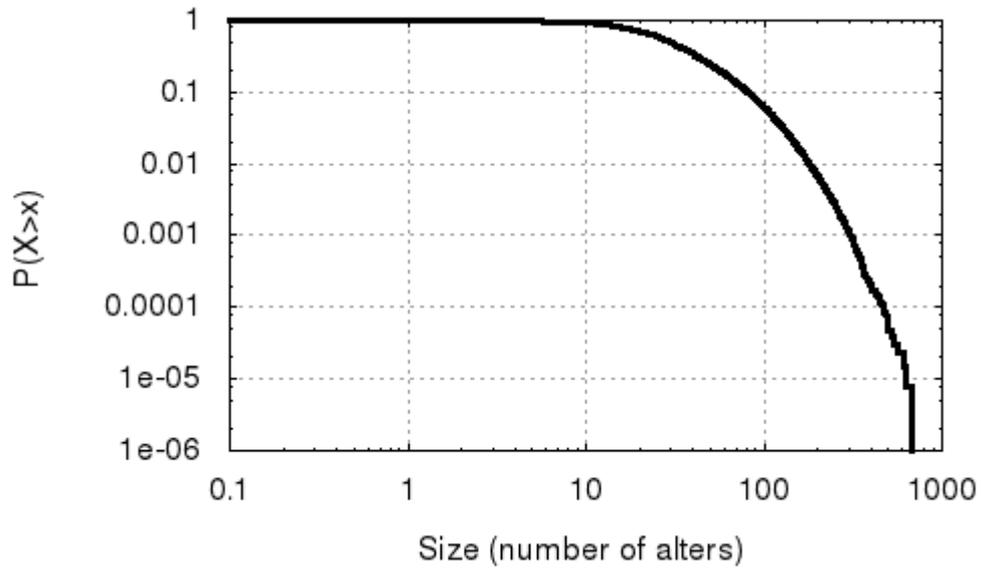

**Figure 3. CCDF of the size of ego networks considering only relationships with contact frequency higher than one message per year (active network) in Facebook dataset #1**

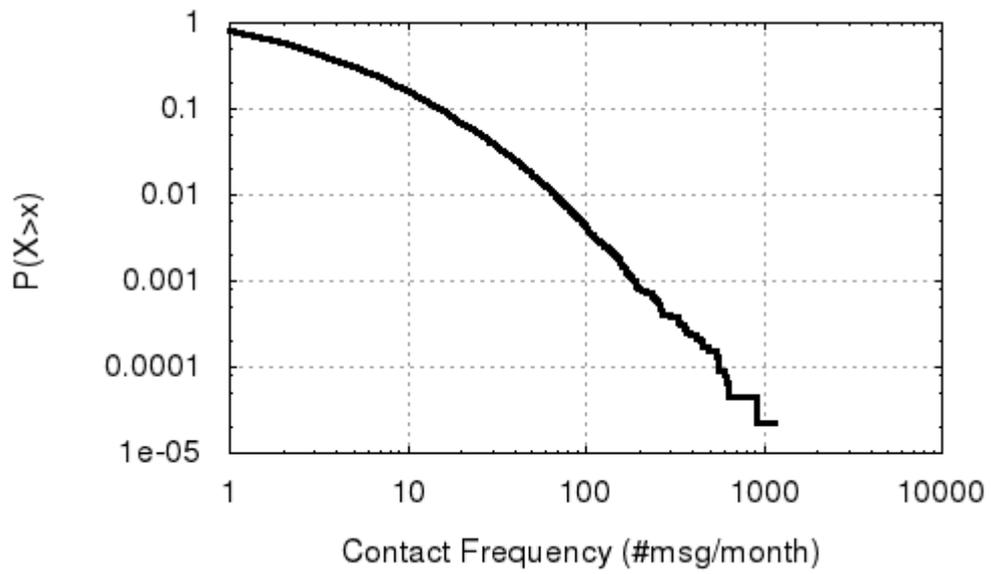

**Figure 4. CCDF of the contact frequency for relationships in Facebook dataset #2**

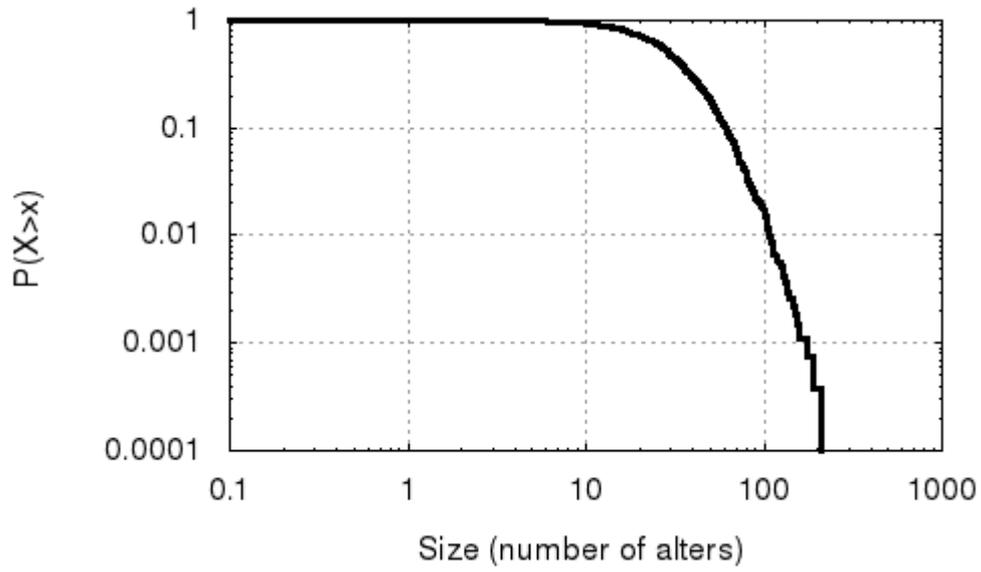

**Figure 5. CCDF of the size of ego networks in Facebook dataset #2**

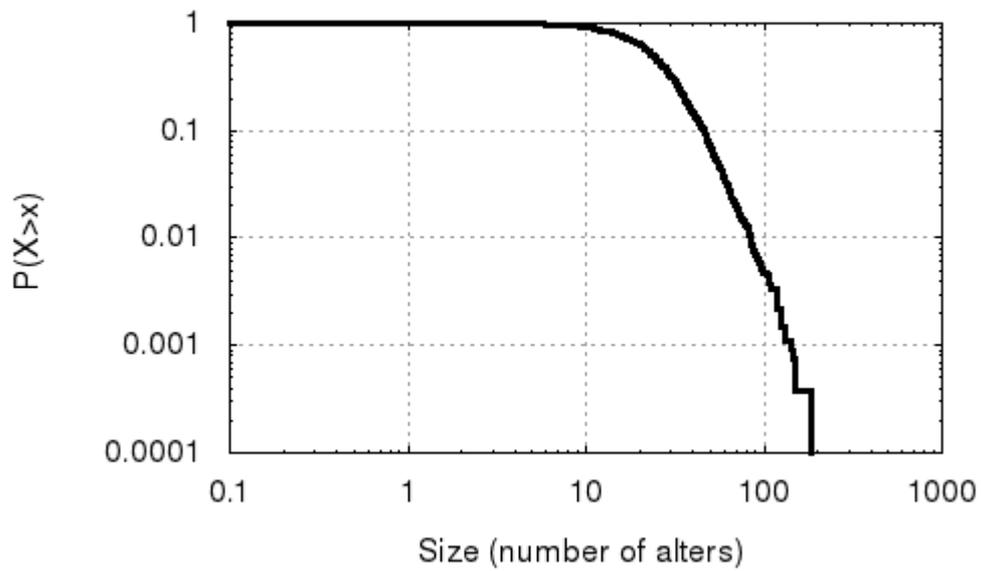

**Figure 6. CCDF of the size of ego networks considering only relationships with contact frequency higher than one message per year (active network) in Facebook dataset #2**

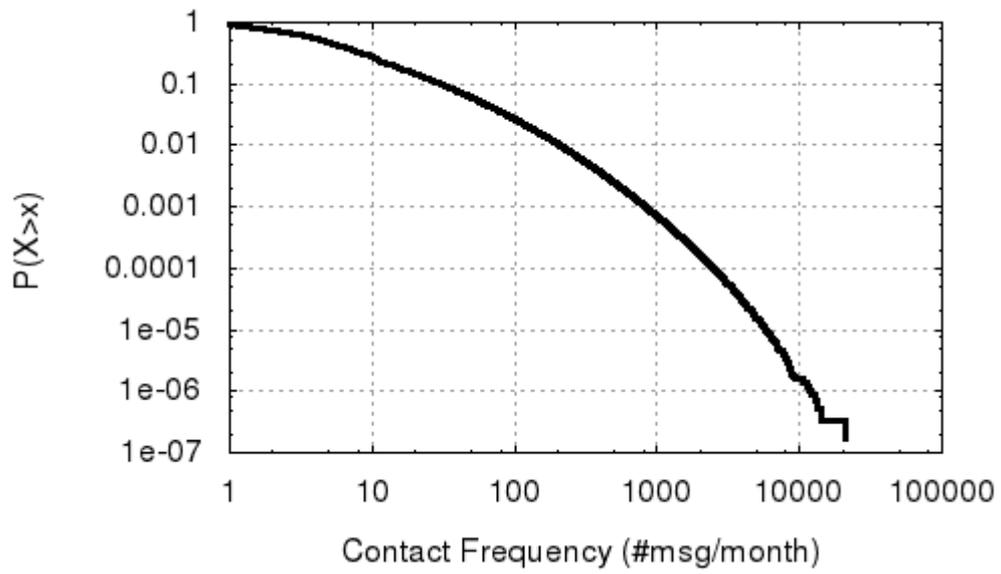

**Figure 7. CCDF of the contact frequency for relationships in Twitter**

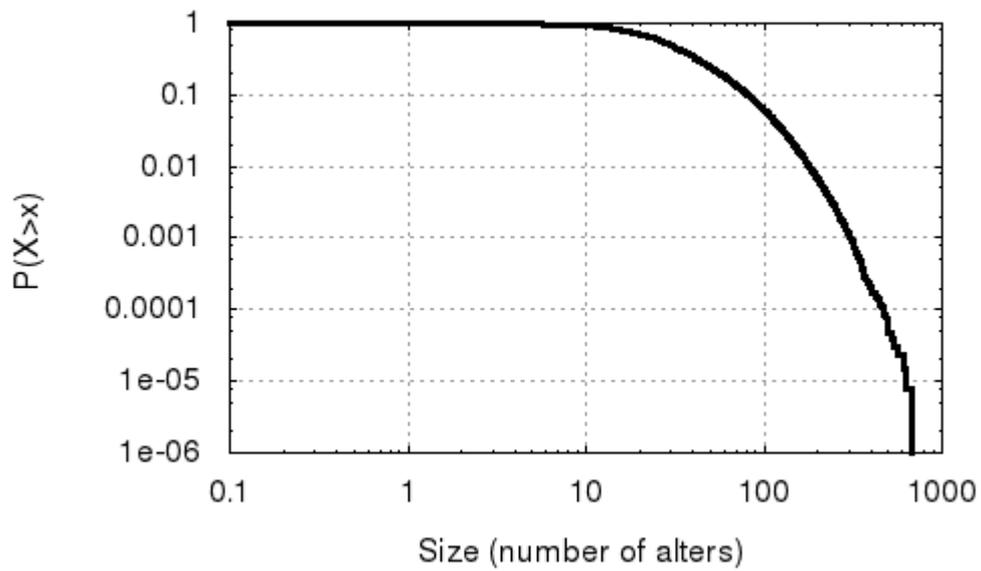

**Figure 8. CCDF of the size of ego networks in Twitter**

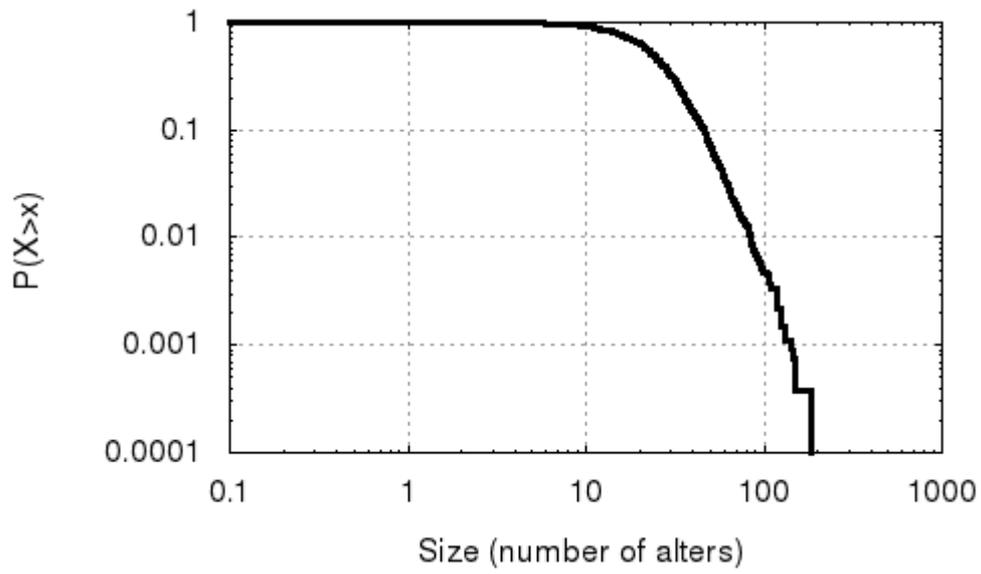

**Figure 9. CCDF of the size of ego networks considering only relationships with contact frequency higher than one message per year (active network) in Facebook dataset #2**

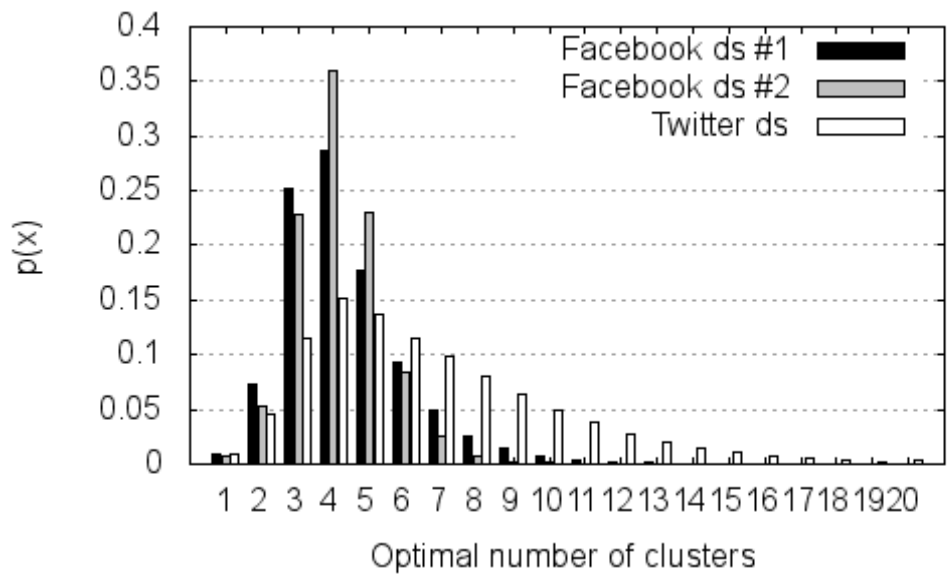

**Figure 10. Results of *k*-means cluster analysis for (a) Facebook dataset #1, (b) Facebook dataset #2 and (c) the Twitter dataset**